\documentclass[12pt,a4paper]{amsart}
\usepackage{amsmath}
\usepackage{amsfonts}
\usepackage{amssymb}
\usepackage{amsthm}
\usepackage{newlfont}
\usepackage{graphicx}
\usepackage{amscd}
\begin{document}
\title[Ito equations out of domino cellular automaton]
{Ito equations out of domino cellular automaton with efficiency parameters}
\author{Zbigniew Czechowski and Mariusz Bia{\l}ecki}
\address{Z. Czechowski, Institute of Geophysics, Polish Academy of Sciences,
ul. Ks. Janusza 64, Warszawa, Poland}
\email{zczech@igf.edu.pl}
\address{M. Bia{\l}ecki, Institute of Geophysics, Polish Academy of Sciences,
ul. Ks. Janusza 64, Warszawa, Poland}
\email{bialecki@igf.edu.pl}
%
\begin{abstract}
Ito equations are derived for simple stochastic cellular automaton with parameters and compared with 
results obtained from the histogram method. Good agreement for various parameters supports wide applicability of the Ito equation as a macroscopic model.

\noindent {keywords}: {stochastic processes, cellular automata, avalanches, discrete solvable models, ultra-discrete equations, time series} 
\end{abstract}

\maketitle



\section{Introduction}
  
The Ito equation describes evolution of a stochastic diffusion Markov process (of order 1). Its parameters can be nonlinear functions of the process. Therefore, the Ito equation is a good candidate for a nonlinear model of phenomenon which manifests non-regular, random behaviour. Quite often observed variables have macroscopic character, and hence the Ito equation can be considered as a macroscopic model of the complex system, in which microscopic collective interactions have been averaged to an adequate form of terms in the equation.

Various complex processes, geophysical, biological, economical etc., in spite of apparent random appearance, 
can lead to some regular behaviours or patterns. Registered time series are investigated by different, less or more sophisticated methods, which help to find these dependencies. Some interesting regularities may constitute of new macroscopic laws of complex systems. 
     
Time series analysis is well developed branch of science. In the linear case, well known procedures were elaborated (ARMA etc.). However, effective methods of construction of nonlinear models from time series data still are not very satisfactory and require further improvements. Using the Ito equation may provide some progress in the field. 
     
First attempts of determination of the Ito equation were proposed by Haken and Borland (Haken 1988, Borland and Haken 1992a, b, 1993, Borland 1996)). Their procedure (SEQUIN) uses the knowledge of some moments of the joined distribution function, which form constraints for the Maximum Information Principle. The SEQUIN method works sufficiently well for the case of a weak noise. However, it was shown (Rozmarynowska 2009) that the procedure fails in the case of strong multiplicative noise, when long-tail distributions appear. 
     
A purely numerical procedure of construction of the Ito equation from time series data was proposed by Siegert et. al. (1998). This direct procedure, based on the histogram of the joined distribution function, always produces an output (i.e. clouds of points) It gives some approximation of terms in the Ito equation. It is effective for strong noise too, but fitting proper functions to scattered clouds of points is a difficult task. Moreover, it should be underlined that this method may be fallacious, when the time series can not be approximated by a diffusion Markov process. 
It is possible to determine the Markov order of a given time series (Racca et al. 2007) but there is no method  
to verify if it is diffusion process or not. The only inspection we may perform is a comparison of the input time series with that generated by the reconstructed Ito equation. 

An adequacy of approximation of a natural phenomena by nonlinear models is of a great importance. 
To touch the topic in the context of the Ito equation we propose as follows.
First, we replace the natural phenomenon by the domino cellular automaton, which can be fully monitored. 
Due to a complete knowledge of the three level hierarchy of the model, relations between the macroscopic (the Ito equation, moment equations) and the microscopic (rules of the automaton) description are clear and understood. 
Then, we construct the Ito equation in two independent ways. The standard histogram method 
is compared with results derived analytically. Derivation of the Ito equation out of a cellular automaton is a crucial step in this approach. To this aim a special stochastic cellular automaton with avalanches was introduced (Bia{\l}ecki and Czechowski 2010b, Czechowski and Bia{\l}ecki 2010a).
The defining rules of the domino automaton were chosen to satisfy two opposite requirements:
complex behavior and relatively simple mathematical structure. Introducing efficiency parameters in the domino automaton leads to a big diversity of states, which covers wide range of avalanche sizes. 
The goal of this paper is an analytical derivation of the Ito equations for a domino cellular automaton with efficiency parameters and a comparison with results of the histogram method. 

In Section \ref{sec:domino} we shortly introduce the domino cellular automaton with efficiency parameters and present 
equations describing its behaviour in equilibrium state. Properties of the automaton for various efficiency 
parameters are analysed. In Section \ref{sec:fluct} we consider fluctuations around the 
equilibrium and propose suitable approximation formulas. In Section \ref{sec:ito} the Ito equation is derived. 
The comparison of histogram procedure results and analytical formulas are presented. We conclude with 
short remarks in Section \ref{sec:concl}.

\section{Domino cellular automaton with efficiency parameters} \label{sec:domino}

\begin{figure}[t]
	\centering
\includegraphics[height=4cm, width=12cm]{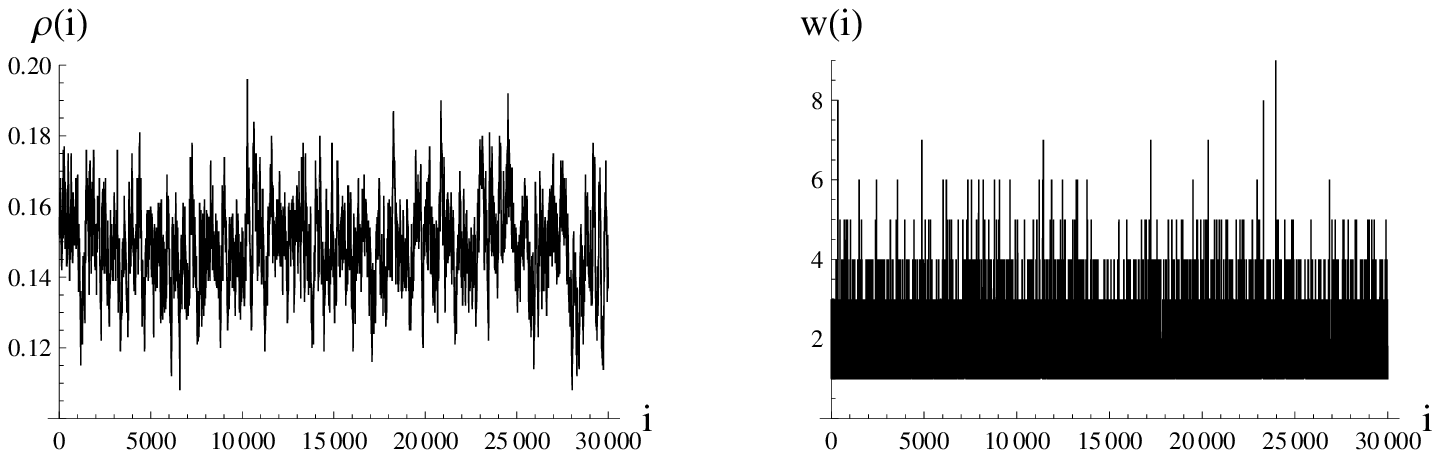} \\
\includegraphics[height=4cm, width=12cm]{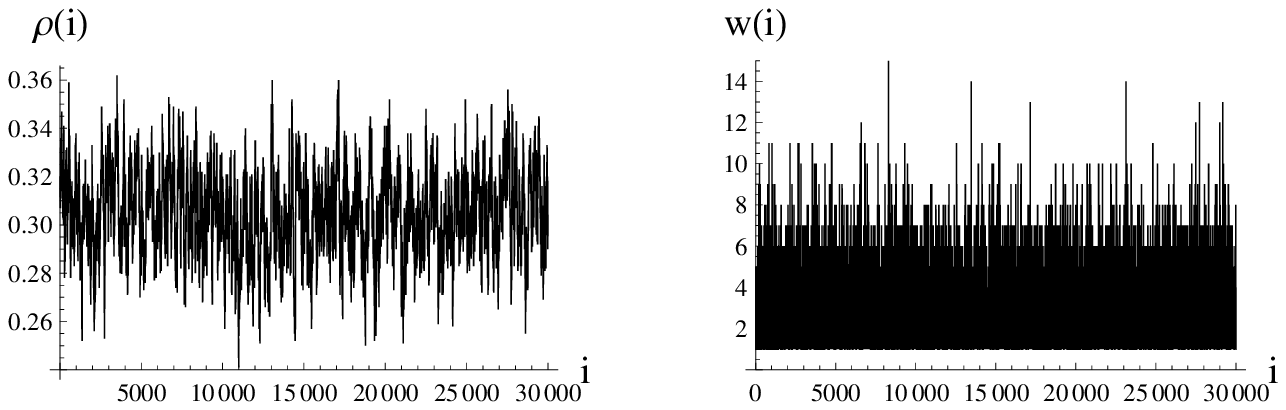} \\
\includegraphics[height=4cm, width=12cm]{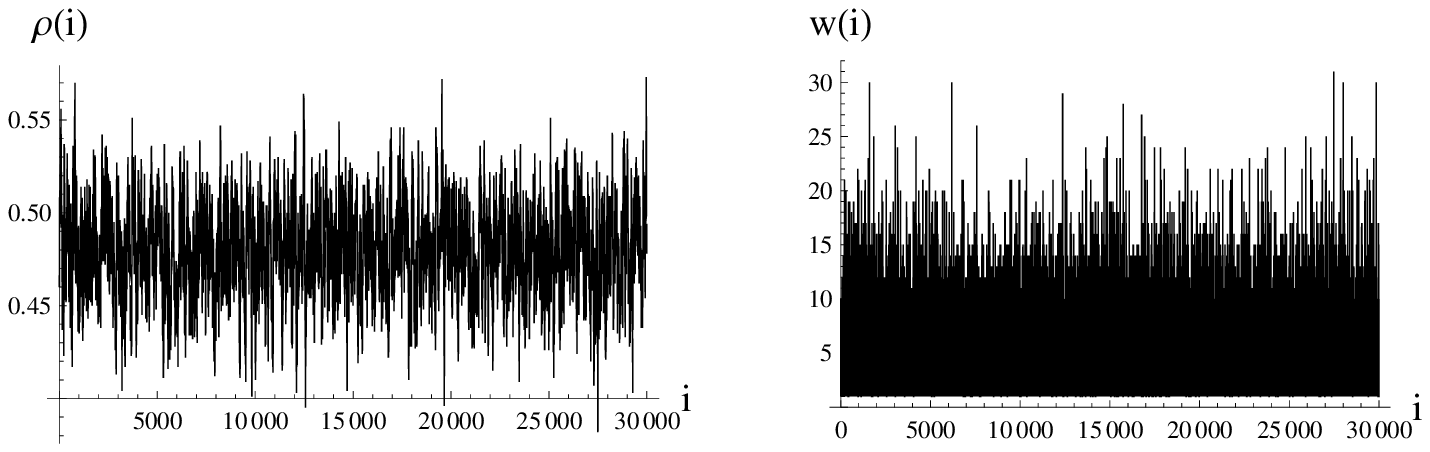}
	\caption{Domino cellular automaton at the quasi-equilibrium state: left –- time series for density $\rho(i)$, right –- time series for avalanche sizes $w(i)$. Three examples for different cases: $\eta = 4$ (upper diagrams), $\eta = 1$ (middle) and $\eta = 0.25$ (lower).}
	\label{fig:Fig1}
\end{figure}

We present shortly rules governing the automaton and some relations derived in Bia{\l}ecki and Czechowski (2010a, b). 
Consider 1-dimensional grid of cells. There are only two states of a cell: empty or occupied (by a single ball).
The evolution rules are as follows. In a time step a single new ball is added to the system to the randomly chosen cell.
\begin{itemize}
\item If the new ball hits an empty cell it becomes occupied with probability $\nu$, or the ball rebounds from the grid with probability $(1-\nu)$. 
\item If the new ball hits an occupied cell it rebounds with probability $(1 - \mu)$, or with probability $\mu$ it knocks out balls from the cell and other adjacent occupied cells from the whole cluster (they become empty –- an avalanche is triggered). 
\end{itemize}

An important macroscopic variable, which describes the state of the automaton is the density $\rho$ i.e. the rate of number of occupied cells to the grid size. We monitor two time series generated by the cellular automaton: the avalanche sizes $w(i)$ and the density $\rho(i)$ computed after each avalanche. Figure \ref{fig:Fig1} presents three examples of such time series for three chosen values of the ratio $\mu/\nu = 4, 1,$ and $0.25$. An increase of the average density, of the average avalanche size and of the range of fluctuations of $\rho$ with decreasing $\mu/\nu$ is evident.

A state of the automaton can be characterized by the cluster size distribution $n_i$, i.e. by numbers of clusters of size $i$ (normalized by the size of the grid).
The following relations, which are exact in the equilibrium state, were derived: 
\begin{eqnarray} 
\quad \quad n_1 &=& \frac{1}{\eta+2} (1-\rho)  - \frac{2}{3} n + \frac{1}{3}n_1^0  \label{eq:n1}\\
\quad \quad n_2 &=& \frac{2}{2\eta+2} \left( 1- \frac{n_1^0}{n} \right) n_1  \label{eq:n2} \\
\quad \quad n_i &=& \frac{1}{\eta i+2} 
\left( 
2 n_{i-1}  \left( 1- \frac{n_1^0}{n} \right) 
+ n_1^0 \sum_{k=1}^{i-2} \frac{n_k n_{i-1-k}}{n^2} 
\right) \quad \text{for} \quad i\geq 3.     \label{eq:ni}
\end{eqnarray}
Here $\eta$ is the rate $\mu / \nu$, $n$ is the number of all clusters, $n= \sum_{i\geq1} n_i$, and
$\rho$ is the density of occupied cells, $\rho= \sum_{i\geq1} i n_i$. The $n_1^0$ is the number of single empty cells 
and is given by the following formula $n_1^0  = \frac{ 2 n}{\left(3 + \frac{2\eta\rho}{n} \right)}.$

The following simple balance equations for moments of $n_i$ were also derived: 
\begin{eqnarray}
2m_0 + (1+\eta) m_1 &=& 1, \label{eq:moments1} \\
 m_1 + \eta m_2  &=& 1, \label{eq:moments2} 
\end{eqnarray}      
where: $m_0 = n$, $m_1 = \rho$ and $m_2=\sum_{i\geq 1} n_i i^2$. The average cluster size $<i>$, the average avalanche size $<w>$ and the average square deviation from the average cluster size  $<T>$ (i.e. the analogue of temperature) are expressed by these moments: 
\begin{eqnarray}
  <i> &=& \frac{m_1}{m_0} =\frac{2\rho}{1-(1+\eta)\rho}\label{eq:avri} \\
  <w> &=& \frac{m_2}{m_1} = \frac{1-\rho}{\eta\rho} \label{eq:avrw} \\
  <T> &=& \sum_{i\geq 1} n_i (i-<i>)^2 = \rho (<w>-<i>)\label{eq:avrT}. 
\end{eqnarray}           
The equilibrium value of the density $\rho_e$ can be found by numerical solution of implicit algebraic equation
\begin{equation}
\rho= \sum_{i\geq1} i n_i,
\label{eq:defrho}
\end{equation}
where  $n_i$  are given by equations \eqref{eq:n1}-\eqref{eq:ni}.  

\begin{table}[t]
	\centering
	\begin{tabular}{c|cccc}
	& \multicolumn{2}{c} {$\rho_e$} & \multicolumn{2}{c} {$<w>$} \\
	&  analytical &  simulation &  analytical &  simulation \\
	\hline
$\eta=4$ &   0.1501 & 0.1498 &  1.4147 & 1.4159\\ 	
$\eta=1$	  &  0.3075 & 0.3068   &  2.2520 & 2.2421 \\
$\eta={1}/{4}$	  &    0.4815   & 0.4801 & 4.3074  & 4.3382 \\
		\end{tabular}
			\vspace{.5cm}
	\caption{Comparison of values of equilibrium density $\rho_e$ and average avalanche size $<w>$ calculated from 
	equations with those obtained from simulations.}
	\label{tab:Table1}
\end{table} 

\begin{figure}[t]
	\centering
	\includegraphics[height=10cm, width=12cm]{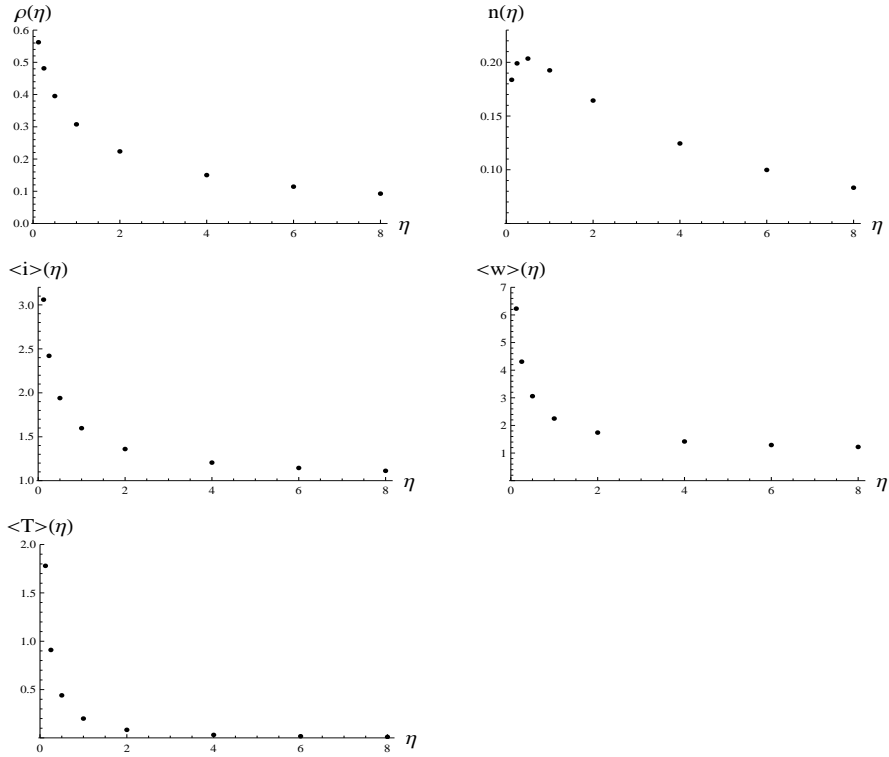}
	\caption{Dependences of equilibrium variables: $\rho_e$, $n$, $<i>$, $<w>$  and  $<T>$ on the ratio of efficiency parameters $\eta =\mu / \nu $.}
	\label{fig:Fig2}
\end{figure}
     
Figure \ref{fig:Fig2} shows dependencies of equilibrium variables: $\rho_e$, $n$, $<i>$, $<w>$  and  $<T>$,  on the ratio $\eta$. As it was expected, they decrease with increasing $\eta$, because greater probability $\mu$ (relevant {\bf} to $\nu$) of triggering the avalanche decrease the number of occupied cells on the grid. Only the total number of clusters $n$, has a maximum for $\eta$ around $1/4$. 
Table \ref{tab:Table1} display a comparison of  values of  $\rho_e$  and  $<w>$ calculated from 
equations with respective values taken from simulation.

\section{Formulas for fluctuations around the equilibrium} \label{sec:fluct}

For deviations from the equilibrium state $\rho_e$ formulas from Section \ref{sec:domino} are not valid. Therefore, we use the approximation introduced in our parallel paper (Czechowski and Bia{\l}ecki, 2010b). It was based on the assumption that the following geometric-like form for $n_i(\rho)$ is maintained:
\begin{eqnarray}
n_1(\rho) &=& (1-\rho)^2 a_1(\rho) = (1-\rho)^2 a_1 \rho, \label{eq:an1}  \\
n_k(\rho) &=& n_{k-1}(\rho)a_k(\rho) \quad \text{for} \quad 4 \geq k \geq 2,  \\
n_k(\rho) &=& n_{k-1}(\rho)(a_5(\rho))^{k-4} \quad \text{for} \quad k \geq 5,  \label{eq:ank}
\end{eqnarray}
where $a_k(\rho)$, $k = 1, \ldots, 5$  are linear functions of $\rho$. The forms of the formulas
 \eqref{eq:an1}-\eqref{eq:ank} extend the 1-D percolation geometric cluster size distribution, $n_k = \rho^k(1-\rho)^2$. Numerical simulations of the domino automaton confirm a validity of this approximation. 

\begin{figure}[t]
	\centering
	\includegraphics[height=3.5cm, width=12cm]{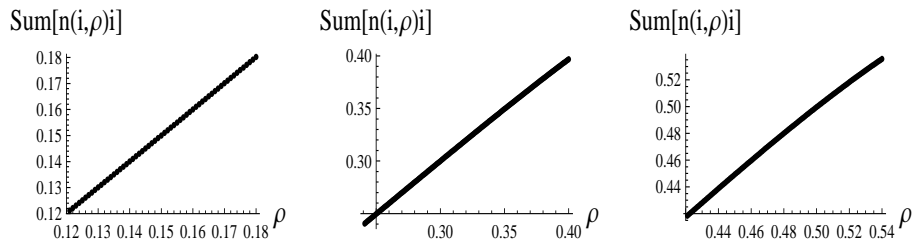} 
	\caption{Sum $\sum_{i\geq1} i n_i$ computed by using  formulas \eqref{eq:an1}-\eqref{eq:ank} versus $\rho$ for three cases: $\eta = 4$ (left graph), $\eta = 1$ (middle), $\eta = 0.25$ (right).}
	\label{fig:Fig3}
\end{figure}

Linear functions $a_k(\rho)$, $k = 2, \ldots, 5$, are found from the Taylor expansion of ratios: 
\begin{equation}
  n_k(\rho)/n_{k-1}(\rho)
\end{equation}                
around $\rho_e$ to the order 1. Using the procedure presented in Czechowski and Bia{\l}ecki (2010b) numerical formulas for $a_k(\rho)$, $k = 1,\ldots, 5$,  for chosen values of the parameter $\eta$ were found and they are presented in Table \ref{tab:Table2}. 
As a check for these expressions, we we plot the first moment of $n_k(\rho)$ (which should be equal to the density $\rho$) as a function of $\rho$ for chosen parameters $\eta$. Figure \ref{fig:Fig3} shows satisfactory fit in the full range of variability of $\rho$ for all three values of $\eta$.  

\begin{table}[t]
	\centering
	\begin{tabular}{c|ccc}
	&  $\eta=4$ &  $\eta=1$ &  $\eta=0.25$ \\
	\hline
$a_1(\rho)$   &   $0.9528 \rho$  					&  $0.8365 \rho$  				 &   $0.7348\rho$ \\
$a_2(\rho)$   &   $0.6449 \rho + 0.0715$  &  $0.7032 \rho + 0.1223$  &   $0.5669\rho + 0.1470$ \\
$a_3(\rho)$   &   $2.4095 \rho - 0.1846$  &  $1.4316 \rho - 0.0484$  &   $1.3742\rho - 0.0546$ \\
$a_4(\rho)$   &   $0.5163 \rho + 0.0989$  &  $0.5651 \rho + 0.2273$  &   $0.4908\rho + 0.3761$ \\
$a_5(\rho)$   &   $0.5139 \rho + 0.0994$  &  $0.5569 \rho + 0.2328$  &   $0.4772\rho + 0.4054$ \\
		\end{tabular}
			\vspace{.5cm}
	\caption{Numerical values of functions $a_k$ for expressions \eqref{eq:an1}-\eqref{eq:ank}.}
	\label{tab:Table2}
\end{table}

\section{Derivation of the Ito equation} \label{sec:ito}
First of all, we should decide which time series, of the density $\rho(i)$ or of the avalanche size $w(i)$, may be treated as a realization of the Markov diffusion process.
Because the avalanche time series is more complex and not Markovian of order 1, we use the time series for density.
An easy check ensures that automaton rules provide the required 1st order Markov property for $\rho(i)$. 
We emphasize that the following convention is assumed: the state of the grid (and the size of the avalanche) is monitored after each avalanche only. In this way we avoid less interesting stairs-like increase of  $\rho(i)$ in periods between avalanches. 
 
In order to derive the Ito equation 
\begin{equation}
d \rho = a (\rho) dt + \sqrt{b(\rho)} dW(t), 
\label{eq:ito}
\end{equation}
we need the transition probability function $P(\rho + \Delta\rho, t+ \Delta t \ | \ \rho,t)$ (see Risken 1996). Functions $a(\rho)$ and $b(\rho)$ are known to be the drift and diffusion coefficients respectively, and $W(t)$ is the Wiener process. Following our parallel paper (Czechowski and Bia{\l}ecki, 2010b, in which the case $\eta= 1$ was considered) the probability of effective gain $EG(k)$ (an increase of $\rho(i)$ by $k$ occupied cells) and effective loss $EL(k)$ are given by the formulas: 
\begin{eqnarray}
EG(k) &\equiv& P \left( \rho_i+ \frac{k}{N}, i+1 | \rho_i,i \right) =  \nonumber \\
 &&= \sum_{s=k+1} (\nu(1-\rho))^s \mu \rho w_{s-k}(\rho) \quad \text{for} \quad k \geq 0, \label{eq:eg}\\ 
EL(k) &\equiv& P \left( \rho_i - \frac{k}{N}, i+1 | \rho_i,i \right) =  \nonumber \\
 &&= \sum_{s=k}   (\nu(1-\rho))^{s-k} \mu \rho w_{s}(\rho) \quad \text{for} \quad k \geq 1, \label{eq:el}
\end{eqnarray}
where $w_s=\frac{n_s s}{\rho}$ is the probability that the occupied cell is a part of the cluster of size $s$ (which also corresponds to the probability of avalanche of size $s$) and  $N$ is the grid size. 

Balls rebounded off the grid do not trigger avalanches, they set the relative efficiency for triggering an avalanche with respect the efficiency of occupation of an empty cell. Hence, for time series analysis, one can consider only 
active (not rebounded) balls introducing appropriate corrective coefficient.
It follows, in expressions \eqref{eq:eg} and \eqref{eq:el} parameters $\mu$ and $\nu$ should be replaced by effective parameters: 
\begin{eqnarray}
 \mu_e &=& \frac{\mu}{\mu\rho+\nu(1-\rho)} = \frac{\eta}{\eta\rho+1-\rho},   \\
 \nu_e &=& \frac{\nu}{\mu\rho+\nu(1-\rho)} = \frac{1}{\eta\rho+1-\rho}.  
\end{eqnarray} 
As a result, probabilities $EG(k)$ and $EL(k)$ satisfy the normalization condition. 

The drift and diffusion forces in the Ito equation correspond (see Risken 1996) to the first and the second moment of the transition probability: 
\begin{eqnarray}
a(\rho) &\propto& \sum_{k\geq1} k \left(  EG(k)-EL(k) \right), \\  
b(\rho) &\propto& \sum_{k\geq1} k^2 \left(  EG(k)+EL(k) \right).
\end{eqnarray}                                                                                         

\begin{figure}[t]
	\centering
	\includegraphics[height=4cm, width=12cm]{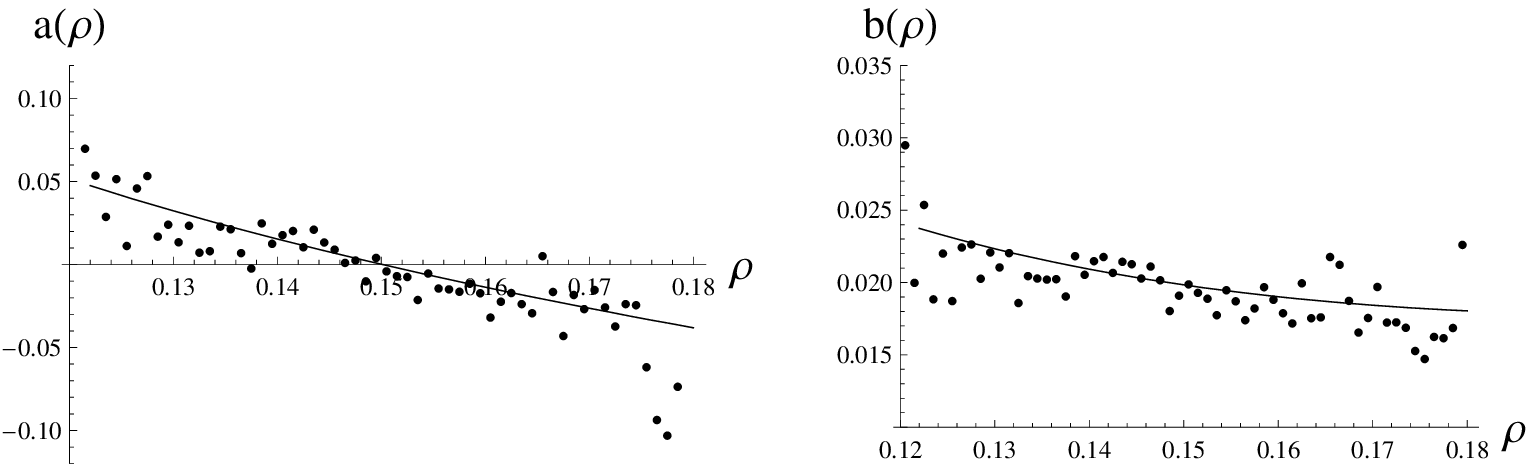} \\
	\includegraphics[height=4cm, width=12cm]{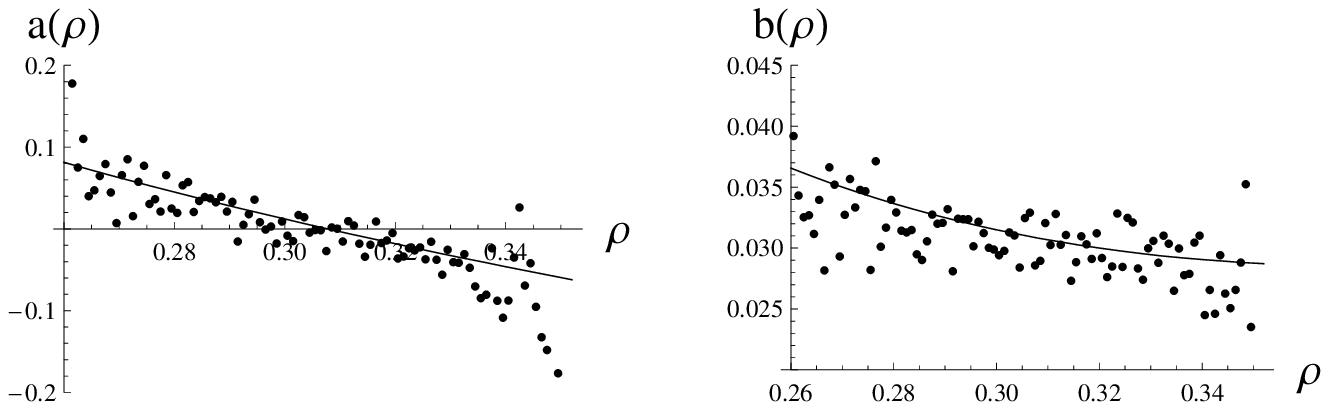} \\
  \includegraphics[height=4cm, width=12cm]{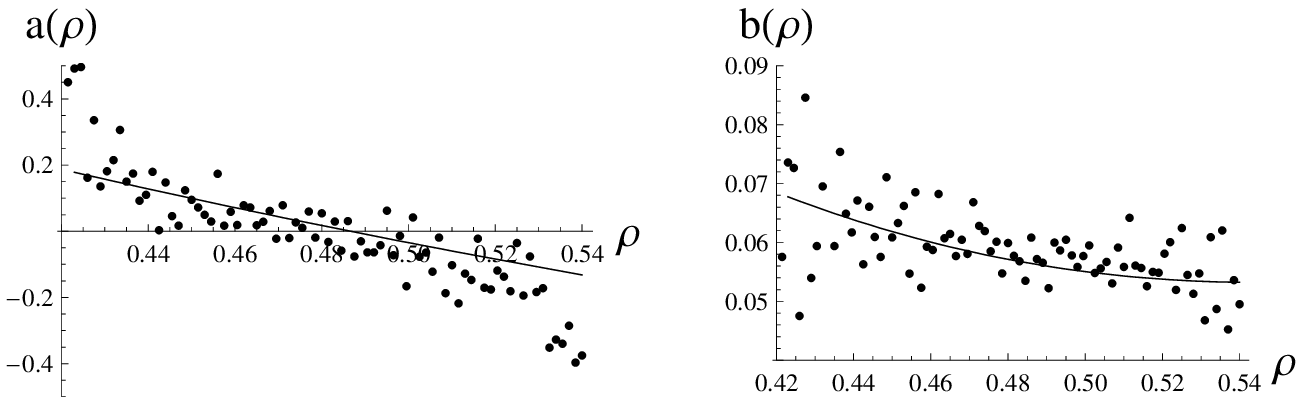}
	\caption{Illustration of functions $a(\rho)$ and $b(\rho)$ in Ito equations for three values of parameters: 
$\eta = 4$ (upper diagrams), $\eta = 1$ (middle), $\eta = 0.25$ (lower). 
Dots represent values reconstructed by the histogram procedure from time series generated by the domino cellular automaton (see Figure \ref{fig:Fig1}), lines –- calculated from formulas \eqref{eq:arhoex} and \eqref{eq:brhoex} using   approximation \eqref{eq:an1}-\eqref{eq:ank} and parameters presented in Table \ref{tab:Table2}.}
\label{fig:Fig4}
\end{figure}

Tedious transformations on sums (see Appendix in Czechowski and Bia{\l}ecki, 2010b) lead to expressions for $a(\rho)$ and $b(\rho)$ in terms of the first, the second and the third moment of  $n_k(\rho)$: 
\begin{eqnarray}
a(\rho)  &\propto& \frac{\nu_e(1-\rho)}{1-\nu_e(1-\rho)} - \frac{1}{\rho} \sum_{k\geq1}  k^2 n_k, \label{eq:arhoex}\\  
b(\rho)  &\propto& \frac{\nu_e(1-\rho)(1+\nu_e(1-\rho))}{(1-\nu_e(1-\rho))^2} +  \label{eq:brhoex}\\
&& -  \frac{2\nu_e(1-\rho)}{\rho(1-\nu_e(1-\rho))} \sum_{k\geq1}  k^2 n_k 
+ \frac{1}{\rho} \sum_{k\geq1}  k^3 n_k. \nonumber
\end{eqnarray}      
Note, the dependence on $\nu_e$ is not essential. The above formulas can be also expressed in terms of $\mu_e$; in fact they depends on $\eta$ and $\rho$ only.                          
The second and the third moments itself can be calculated using approximate formulas \eqref{eq:an1}-\eqref{eq:ank} from Section \ref{sec:fluct}.

\section{Conclusions} \label{sec:concl}

Introduction of efficiency parameters $\mu$ and $\nu$ essentially enriched the automaton. By changing the ratio $\eta=\mu/\nu$ the domino cellular automaton can attain a wide range of values for important variables:  $\rho_e$, $n$, $<i>$, $<w>$  and  $<T>$ (see Figure \ref{fig:Fig2}), which characterize a macroscopic behaviour.  The time evolution of the model can be described by the Ito equation, therefore, we derive analytically a satisfactory approximation of  Ito terms $a(\rho)$ and $b(\rho)$ as functions of the ratio $\eta=\mu/\nu$. For three chosen values of  $\eta$ analytical and simulation results were compared. Figure \ref{fig:Fig4} shows a good fit for these three cases. Thus, we have a simple, fully elaborated model which can manifest a wide range of behaviours. 

     Moreover, we show that the histogram method of reconstruction of Ito equation from time series works well in the case of domino cellular automaton. The automaton plays a role of a complex system, in which energy is slowly accumulated and rapidly released in avalanches. Therefore, this is possible that the histogram procedure can offer an adequate method of nonlinear modelling of similar natural phenomena basing on the time series data. 
The method  fills a gap between linear stochastic models (ARMA etc.) and nonlinear deterministic models (Takens method, see Takens 1981) because it is both stochastic and nonlinear.

\section*{Acknowledgement}

\section*{References}

{\tiny

Bia{\l}ecki M. and Czechowski Z. (2010a)
{\it Analytic approach to stochastic cellular automata: exponential and inverse power distributions out of Random Domino Automaton}, arXiv:1009.4609 [nlin.CG]

Bia{\l}ecki M. and Czechowski Z. (2010b) 
{\it On a simple stochastic cellular automaton with \newline
avalanches: simulation and analytical results},
Chapter 5 in V.~De Rubeis, Z.~Czechowski, and R.~Teisseyre, editors, 
{\em Synchronization and triggering: from fracture to earthquake processes}, Springer, pages 63--75.

Borland L. (1996), 
{\it Simultaneous modelling of nonlinear deterministic and stochastic dynamics}, Physica D, 90, 175-190.

Borland L. and H. Haken (1992a), {\it Unbiased determination of forces causing observed processes. The case of additive and weak multiplicative noise}, Z. Phys. B, 88, 95-103.

Borland L. and H. Haken (1992b), {\it Unbiased estimate of forces from measured correlation funtions, including the case of strong multiplicative noise}, Ann. Physik, 1, 452-460.

Borland L. and H. Haken (1993), {\it On the constraints necessary for macroscopic prediction of stochastkic time-dependent processes}, Rep. Math. Phys., 33, 35-42.

Czechowski Z. and M. Bia{\l}ecki (2010a), {\it Ito equations as macroscopic stochastic models of geophysical phenomena - construction of the models on a base of time series and analytical derivation}, Chapter 6 in V. De Rubeis, Z. Czechowski and R. Teisseyre (Eds.) {\em Synchronization and triggering: from fracture to earthquake processes}, Springer, pages 77-96. 

Czechowski Z. and Bia{\l}ecki M. (2010b), {\it Three-level description of the domino cellular automaton},
arXiv:1012.5902 [nlin.CG]

Haken H. (1988), {\it Information and Self-Organization}, Springer.

Racca E., Laio F., Poggi D. anf Ridolfi L. (2007), {\it Test to determine the Markov order of time series}, 
Phys. Rev. E 75, 011126-1-6. 

Risken H. (1996), {\it The Fokker-Planck Equation}, Springer. 

Rozmarynowska A. (2009) {\it On the reconstruction of Ito models on the base of time series with long-tail distributions}, Acta Geophysica, 57, no 2, 311-329.

Siegert S., Friedrich R. and Peinke J. (1998) {\it Analysis of data sets of stochastic systems}, Phys. Lett. A, 243, 275-280.

Takens F. (1981), {\it Detecting strange attractors in turbulence}, In D.A. Rand and L.S. Young (eds.) {\em Dynamical Systems and Turbulence},  Lecture Notes in Mathematics,  vol. 898, Springer.

}

\end{document}